\documentclass[
aps,
pra,
reprint,
superscriptaddress,
showpacs,
showkeys,
nofootinbib,
floatfix
]{revtex4-1}

\usepackage[utf8]{inputenc}
\usepackage[english]{babel}
\usepackage{amsmath}
\usepackage{amsfonts}
\usepackage{amssymb}
\usepackage{graphicx}

\usepackage{color} 

\usepackage{graphicx}

\usepackage{amsmath}
\usepackage{amsfonts}
\usepackage{amssymb}
\usepackage{amsthm}

\usepackage{url}

\newcommand{\bfA}{{\mathbf{A}}}

\newcommand{\bfE}{{\mathbf{E}}}

\newcommand{\bfH}{{\mathbf{H}}}

\newcommand{\bfr}{{\mathbf{r}}}

\newcommand{\rmd}{{\mathrm d}}

\newcommand{\rme}{{\mathrm e}}

\newcommand{\rmi}{{\mathrm i}}

\newcommand{\ZZ}{{\mathbb Z}}
\newcommand{\sA}{{\mathsf A}}
\newcommand{\bsA}{{\boldsymbol{\mathsf{A}}}}

\newcommand{\sE}{{\mathsf E}}
\newcommand{\bsE}{{\boldsymbol{\mathsf{E}}}}

\newcommand{\sH}{{\mathsf H}}
\newcommand{\bsH}{{\boldsymbol{\mathsf{H}}}}
\newcommand{\sI}{{\mathsf I}}

\newcommand{\sV}{{\mathsf V}}

\newcommand{\derivep}[2]{ \frac{\partial #1}{\partial #2} }

\usepackage{fancyhdr}
\begin{document}

\title{\textsc{DIMOHA}: a time-domain algorithm for traveling-wave tube simulations}

\author{Damien~F.~G.~Minenna}%
 \email[Electronic address: ]{damien.minenna@univ-amu.fr}
\affiliation{%
Centre National d'{\'E}tudes Spatiales, Toulouse, France
}%
\affiliation{%
Aix-Marseille Universit{\'e}, CNRS, PIIM, UMR 7345, 13397, Marseille, France
}%
\affiliation{%
Thales AVS, V{\'e}lizy--Villacoublay, France
}%
\author{Yves~Elskens}%
\affiliation{%
Aix-Marseille Universit{\'e}, CNRS, PIIM, UMR 7345, 13397, Marseille, France
}%
\author{Fr{\'e}d{\'e}ric~Andr{\'e}}%
\affiliation{%
Thales AVS, V{\'e}lizy--Villacoublay, France
}%
\author{Alexandre~Poy{\'e}}%
\affiliation{%
Aix-Marseille Universit{\'e}, CNRS, PIIM, UMR 7345, 13397, Marseille, France
}%
\author{J{\'e}r{\^o}me~Puech}%
\affiliation{%
Centre National d'{\'E}tudes Spatiales, Toulouse, France
}%
\author{Fabrice~Doveil}%
\affiliation{%
Aix-Marseille Universit{\'e}, CNRS, PIIM, UMR 7345, 13397, Marseille, France
}%

\date{Received May 29, 2019; revised July 5, 2019; accepted
July 10, 2019. Date of publication July 31, 2019 \\ date of current version
August 21, 2019}

\begin{abstract}
To simulate traveling-wave tubes (TWTs) in time domain and more generally the wave-particle interaction in vacuum devices, we developed the DIscrete MOdel with HAmiltonian approach (\textsc{dimoha}) as an alternative to current particle-in-cell (PIC) and frequency approaches. 
Indeed, it is based on a longitudinal $N$-body Hamiltonian approach satisfying Maxwell's equations. 
Advantages of \textsc{dimoha} comprise: 
(i)~it allows arbitrary waveform (not just field envelope), including continuous waveform (CW), multiple carriers or digital modulations (shift keying);
(ii)~the algorithm is much faster than PIC codes thanks to a field discretization allowing a drastic degree-of-freedom reduction, along with a robust symplectic integrator;
(iii)~it supports any periodic slow-wave structure design such as helix or folded waveguides; 
(iv)~it reproduces harmonic generation, reflection, oscillation and distortion phenomena; 
(v)~it handles nonlinear dynamics, including intermodulations, trapping and chaos.
\textsc{dimoha} accuracy is assessed by comparing it against measurements from a commercial Ku-band tapered helix TWT and against simulations from a sub-THz folded waveguide TWT with a staggered double-grating slow-wave structure. 
The algorithm is also tested for multiple-carriers simulations with success. 
\end{abstract}

\keywords{Traveling-wave tubes (TWTs), DIMOHA, simulation, AM/AM, AM/PM, nonlinear signals, time domain, wave-particle interaction, helix, folded waveguide, gratings, slow-wave structure, Hamiltonian, N-body dynamics, communication systems, vacuum electronics. \\
$~$ \\
\bigskip
Published as: IEEE Trans.\ Elec.\ Devices, \textbf{66}(9): 4042-4047 (2019), doi: 10.1109/TED.2019.2928450. }


\maketitle
\thispagestyle{fancy}

\section{Introduction}

Nowadays, simulating traveling-wave tubes \cite{min19epjh} (TWTs) is still a challenge due to the large number of parameters involved and because, in nonlinear regime, the large power of high frequency TWTs and their broad band spectrum generate critical instabilities.
The physical process at the heart of the device is the wave-particle interaction, whereby electrons transfer their momentum to amplify telecommunication signals, as, e.g., an inverse Landau damping: synchronization \cite{dov05} occurs when the phase velocity of the wave (determined by the tube geometry) is close to the particles speed.

Currently, to model the propagating wave interacting with the beam, two kinds of numerical approaches are used: time domain and frequency (steady-state) domain modeling.
The first kind allows, in principle, to investigate the majority of TWT problems such as reflections, oscillations, harmonic and intermodulation generations, multi-frequency signals. 
But its main drawbacks are the computation size and duration. This is especially true when studying oscillations and intermodulations that require strong resolution in time to be accurate.
Generally, time domain codes are based on three-dimensional (3D) meshes (thousands to millions of points) based on Maxwell's equations such as particle-in-cell (PIC) codes relying on kinetic (e.g.\ Vlasov) equations (like \textsc{cst} \cite{saf18,cst}, \textsc{karat} \cite{karat} or \textsc{mafia} \cite{wei97}). 
To reduce the computational costs, one can use specialized frequency (harmonic or steady-state) models (like \textsc{mvtrad} \cite{wal99}, \textsc{christine} \cite{ant97} or \textsc{bwis} \cite{li09}) which rely on reduction models such as equivalent circuits envelope models.

We propose a third option, using an $N$-body (a.k.a.\ many body) description \cite{min18,min17,min19ivec} to design a specialized theory in time domain and large-signal regime as an alternative to both PIC and frequency approaches.
By nature, this description is deemed slower and computationally far more expensive than PIC since all particles should be considered.
For periodic structures, we overcome this slowness thanks to a drastic degrees-of-freedom (dof) reduction \cite{and13,min18,min19ivec,min17} (Kuznetsov's discrete model).
The reduction is extreme ($\sim 100 \, 000$ dofs for space TWTs) compared to finite difference techniques in PIC codes that often involve several millions of dofs to reach the same accuracy. 
Moreover, our approach is one-dimensional (1D), while generally time domain codes require simulating the whole slow-wave structure (SWS) of the tube in 3D. Therefore, the model consistency and conservation properties are more easily achieved.
This model reduction is combined with the $N$-body hamiltonian approach \cite{min18,min17} leading to a better control on conservation properties, including Poincar{\'e}-Cartan invariants (dynamics geometry) \cite{arn89}. From this, we built a symplectic \cite{hlw10,sai01} algorithm \textsc{dimoha} (DIscrete MOdel with HAmiltonian approach), enabling us to increase the numerical time step without incurring too much error on results.
To assess its validity, we compare this algorithm with measurements from a commercial 140W Ku-band tapered helix TWT and with simulations from a sub-THz staggered TWT.

This paper is dedicated to presenting our algorithm, its capacities and some results obtained with it. However, to keep the paper short and focus on results, we defer the overview of the underlying theory to a forthcoming report.

\section{The algorithm}

\subsection{Algorithm description}

The current version of \textsc{dimoha} is developed jointly between Aix-Marseille Universit{\'e}--CNRS, Thales AVS/MIS and the Centre National d'{\'E}tudes Spatiales in France.
It  is written in modern Fortran and parallelised with the Message Passing Intergace (MPI) librairies.
\textsc{dimoha} simulates, in 1D, the interaction between electrons and waves in the slow-wave structure (SWS) of industrial TWTs.

\begin{figure}[!t]
\centering
\includegraphics[width=\columnwidth]{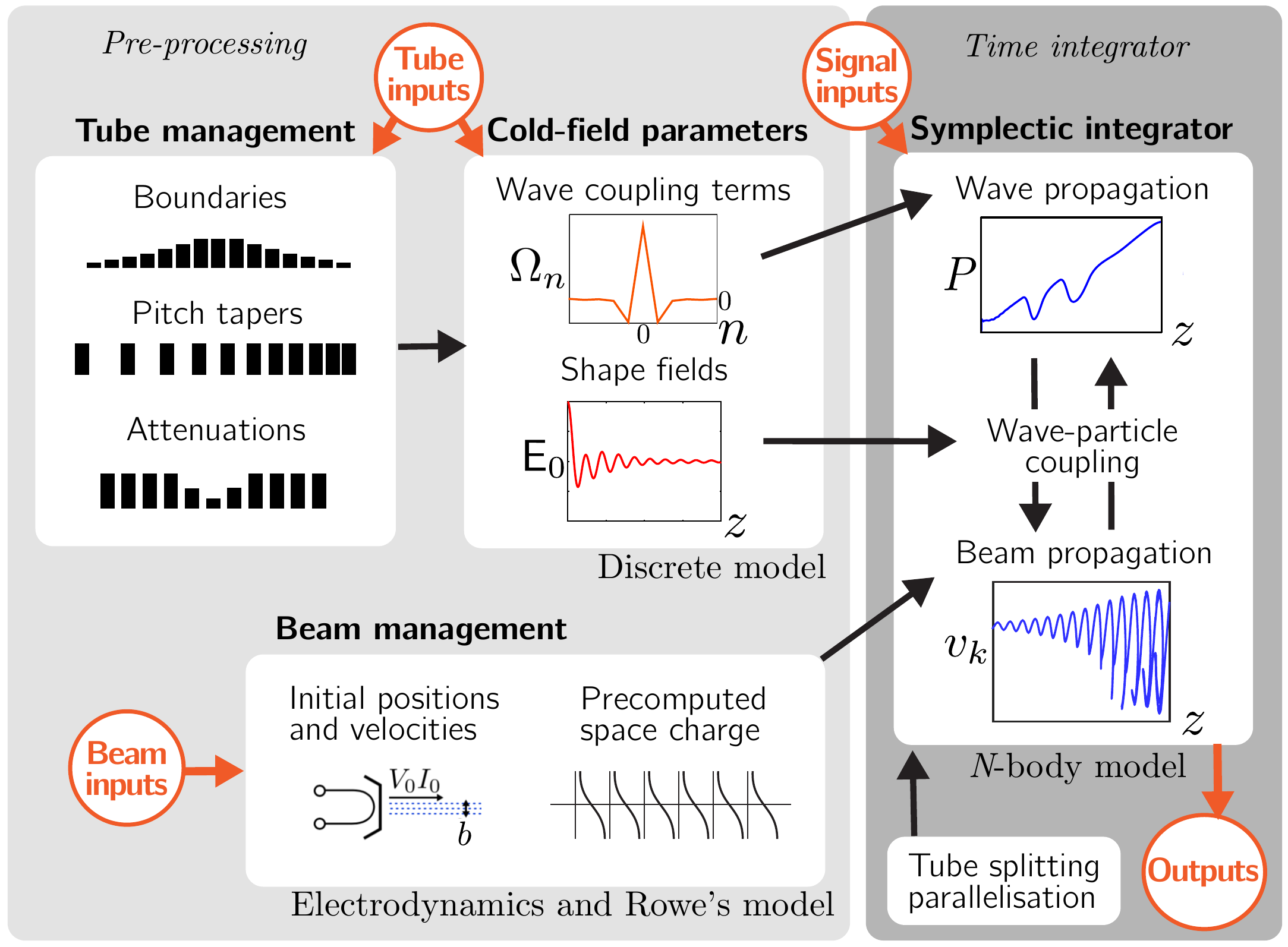}
\caption{Process workflow of \textsc{dimoha}. The algorithm is divided in three parts: (i) the pre-processing to compute cold-field parameters; (ii) the time integrator; (iii) the post-processing (not shown) that performs diagnostics and computes the RF power. Inputs and outputs are listed in Table~\ref{t:inout}.}
\label{f:dimoha}
\end{figure}

In time domain, a comparison with \textsc{mvtrad} \cite{wal99} (a 2.5D frequency-domain (field) particle-in-cell (beam) code specialized for helix TWT) has already shown \cite{min18} perfect agreement at power saturation, in nonlinear regime (trapping), in the middle of the transmission band of a basic tube without taper nor attenuators, using a preliminary version of our algorithm.
The novelty, here, is that we adapted the model to take into account non-periodic structure (taper) and attenuators that are major components of industrial TWTs.
The main vocation of \textsc{dimoha} is to provide a complete tool that can be used for industrial design and research activities as well as for device assessement by telecom operators.

As shown on Fig.~\ref{f:dimoha}, \textsc{dimoha} is composed of three parts (see Sections~\ref{s:CF}-\ref{s:IntDiag}). The first part (pre-processing) computes the cold-field parameters from the tube geometry. The second part (time integrator) is the heart of the algorithm that simulates the wave-electron interaction in the slow-wave structure up to a given time $t_{\rm max}$.
The last part (post-processing) performs final diagnostics after the interaction.
The parametric inputs are given by Table~\ref{t:inout}. In continuous wave regime, only the frequency and power of the injected wave are required.

\begin{table}[t]
\centering
\caption{\label{t:inout} Inputs and outputs of \textsc{dimoha}}
\begin{footnotesize}
\begin{tabular}{rl}
\hline
\multicolumn{2}{c}{\textbf{Inputs}}                                                \\ \hline
Tube : & Pitch $d$, dispersion relation $\omega(\beta)$, \\
 (per cell) & impedance $Z_{\rm c} (\beta)$ and attenuations $\alpha_n$\\
 & \\
Beam : & Current $I_0$  and potential $V_0$  \\
(cathode) & beam diameter $b$  \\
 & \\
Signal: & Injected 1D electric field in time at the tube inlet            \\  \hline
  \hline
\multicolumn{2}{c}{\textbf{Simulation parameters}}    \\
  \hline
  $t_{\rm max}$ & Duration of the interaction \\ 
  $\Delta t$ & Time step size \\
  $\delta$ & Initial spacing between macro-electrons \\
  $N_{\rm os}$  &  Number of mesh points per cell \\
  $N_{\rm ph}$ & Range of coupling between cells \\ \hline \hline
\multicolumn{2}{c}{\textbf{Outputs}}                                               \\ \hline
\multicolumn{2}{l}{1D electric field in time inside the tube}                     \\
\multicolumn{2}{l}{1D distribution function of electrons (positions, velocities)}
\end{tabular}
\end{footnotesize}
\end{table}

\subsection{The discrete model}

 Our algorithm \textsc{dimoha} is based on the Kuznetsov discrete model \cite{kuz80,rys07,rys09,ber11,ber11b,and13,ter17,min18,min19pc} which uses an exact field decomposition for periodic structures for large-signal regimes.
The SWS is directed along the $z$-axis with pitch $d$ and cell index $n \in \ZZ$. We assume the fundamental mode of propagation is dominant, so we neglect other modes. The scalar potential $\phi_{\rm sc}$ of the beam space charge satisfies Poisson's equation.
Within the discrete model, the RF electric (circuit part) and magnetic fields in the whole structure read
\begin{align}
\bfE(\bfr,t) &=  \sum_{n \in \ZZ} \sV_{n}(t) \bsE_{-n}(\bfr)  \, , \label{e:FieldE} \\
\bfH(\bfr,t) &=  \sum_{n \in \ZZ} \rmi \sI_{n}(t) \bsH_{-n}(\bfr) \, , \label{e:FieldH}
\end{align}
that satisfy Maxwell's equations.
$\sV_{n},\sI_{n}$ are the cell-discretized temporal field amplitudes, and $\bsE_{n},\bsH_{n}$  are the ``shape'' functions of the fields depending only on the structure geometry (see Eq.~\eqref{e:Ebeta1D}).
The imaginary number ($\rmi$) and the sign in $-n$ are technical conventions \cite{and13,min18} allowing terms $\sV_{n},\sI_{n},\bsE_{n}, \rmi \bsH_{n}$ to be real.
To compute time variables $\sV_{n},\sI_{n}$, we need an interaction model like the $N$-body description \cite{min18} (used below) or a fluid model \cite{min19pc}.
Since $\nabla \times \bfA = \mu_0 \bfH$, we also express the vector potential $\bfA =  \sum_{n} \rmi \sI_{n} \bsA_{-n}$ with the same $\sI_{n}$ as for the magnetic field \eqref{e:FieldH}.

Equations~\eqref{e:FieldE}-\eqref{e:FieldH} highlight the interest of the discrete model for dof reduction. Indeed, the time-independent functions $\bsE_{n},\bsH_{n}$ are given by the SWS geometry. 
For a space TWT comprising $n_{\rm max} = 200$ cells, there are only 200 variables $\sV_{n},\sI_{n}$, so that the field involves only 200 dofs.

Beside the time domain approach, one can use a frequency domain version of Kuznetsov's discrete model \cite{rys07,min19pc,ter17,rys} in small signal regime. Its validity was assessed \cite{min19pc} by comparing it with Pierce's four-waves theory \cite{pie50}.

\section{Initialisation of cold-field parameters}
\label{s:CF}
 The first part of \textsc{dimoha} computes cold-field (time-independent) parameters from tube and beam inputs in Table~\ref{t:inout}. Those cold-field parameters, depending on the SWS, can be obtained with a numerical solver or measured.

 \subsection{Coupling terms and attenuators}
 \label{s:couplingterms}

Using the decomposition \eqref{e:FieldE}-\eqref{e:FieldH}, we rewrite sourceless Maxwell equations (equivalently Hamilton equations in our formulation \cite{min18}) as 
 \begin{align}
\frac{{\rm d} \sV_{n}}{{\rm d} t} = - \sum_{n'} \Omega_{n-n'} \sI_{n'}(t) \, , \label{e:evo1} \\
\frac{{\rm d} \sI_{n}}{{\rm d} t} =  \sum_{n'} \Omega_{n-n'} \sV_{n'}(t) \, . \label{e:evo2} 
\end{align}
Those equations represent the wave propagation along the SWS behaving as a harmonic oscillators chain.
Coupling terms $\Omega_{n-n'}$ between cells $n$ and $n'$ are obtained from the Fourier series of the dispersion relation $\omega (\beta) = \sum^{N_{\rm ph}}_{n = -N_{\rm ph}}  \Omega_{n} \, \rme^{\rmi n \beta d}$, with $\beta$ the wavenumber (propagation constant). 
We take the range of coupling $N_{\rm ph}$, estimated in \cite{ber11}, at 5 for helix TWTs and 1 for folded waveguides. 
To take into account any pitch tapering, coupling terms are recomputed at each different cell.
Note that, to compute accurately $\Omega_{n}$, the discrete model requires a broader dispersion relation (e.g. all harmonics) than just the data for the transmission band.
This is an advantage since the wave propagation takes into account the whole dispersion relation and not a mere single point at the frequency of interest as for envelope models. 

To consider attenuators \cite{ber11,ber11b}, we add $- \alpha_n \sV_{n}$ to the right side of Eq.~\eqref{e:evo1}. 
\textsc{dimoha} can process spatially distributed attenuations in tubes. We can either provide in input the coefficients in Np/m or dB$/$wavelength from the tube permittivity or provide the total transmission loss $S_{12}$ from measurements and estimate the average coefficients.
For tubes with severs and local attenuators, the $\alpha_n$'s are adjusted consequently.

 \subsection{Tube's meshes}
 
The simulation uses two meshes. The position of each cell, spaced by  pitch $d$, is given by $z_{\rm cell}$. To reproduce accurately the space charge field and the coupling between the field and particles, an adapted mesh $z_{\rm field}$ oversamples the basic mesh by $N_{\rm os}$ points per cell. For TWTs, $N_{\rm os}$ does not need to be large (e.g.\ $N_{\rm os} = 20$ in our simulations).
 
Without beam, Eqs~\eqref{e:evo1}-\eqref{e:evo2} can be seen as the equations of a harmonic oscillators chain, so an abrupt stop of the SWS at the edge of the tube would lead to a total reflection of the signal. To avoid this, we lengthen the tube and use another coefficient $\alpha_n$ that slowly increases when moving away from the physical part (perfectly matched layer (PML) method).

\begin{figure}[!t]
\centering
a) \includegraphics[width=0.9\columnwidth]{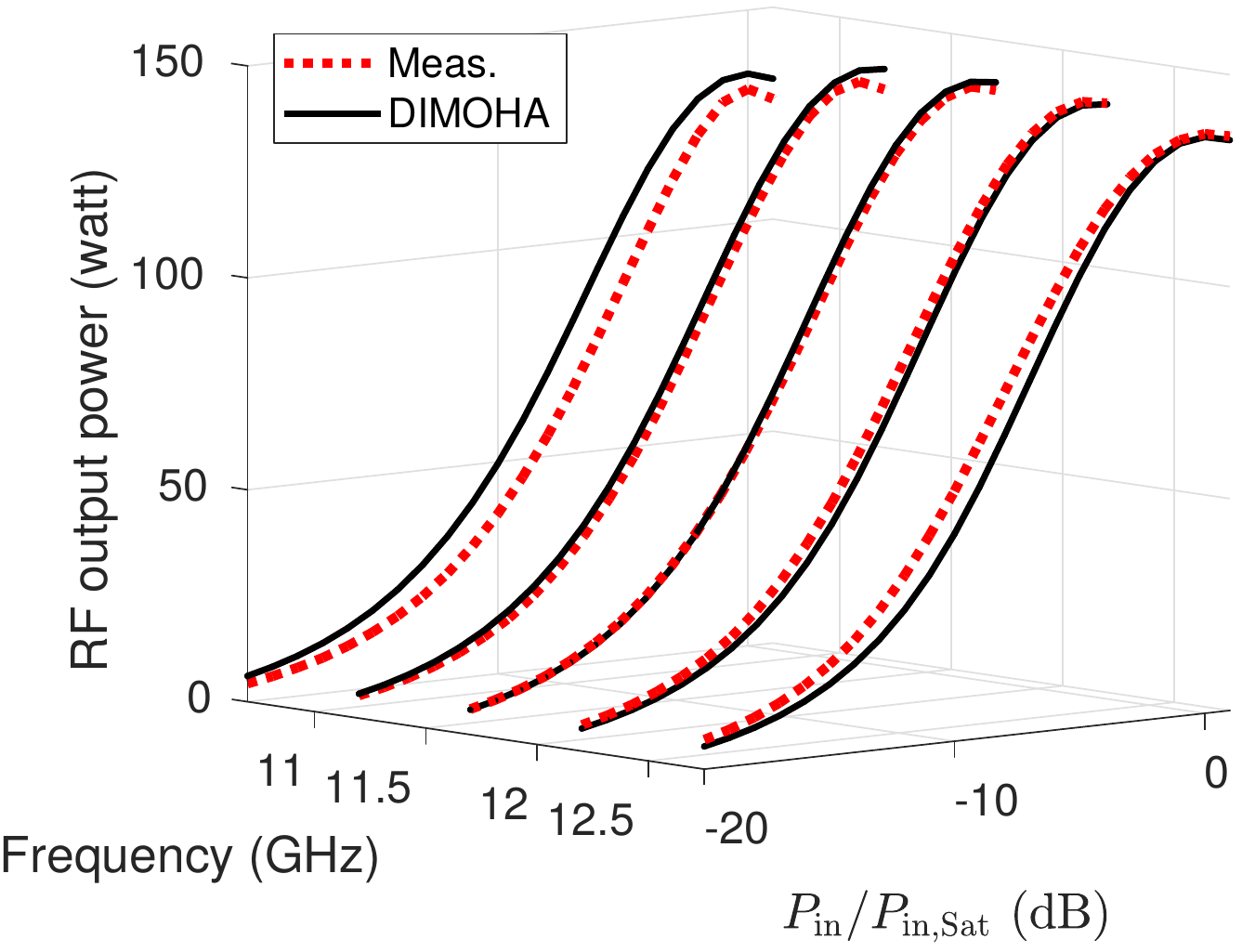} \\
b) \includegraphics[width=0.9\columnwidth]{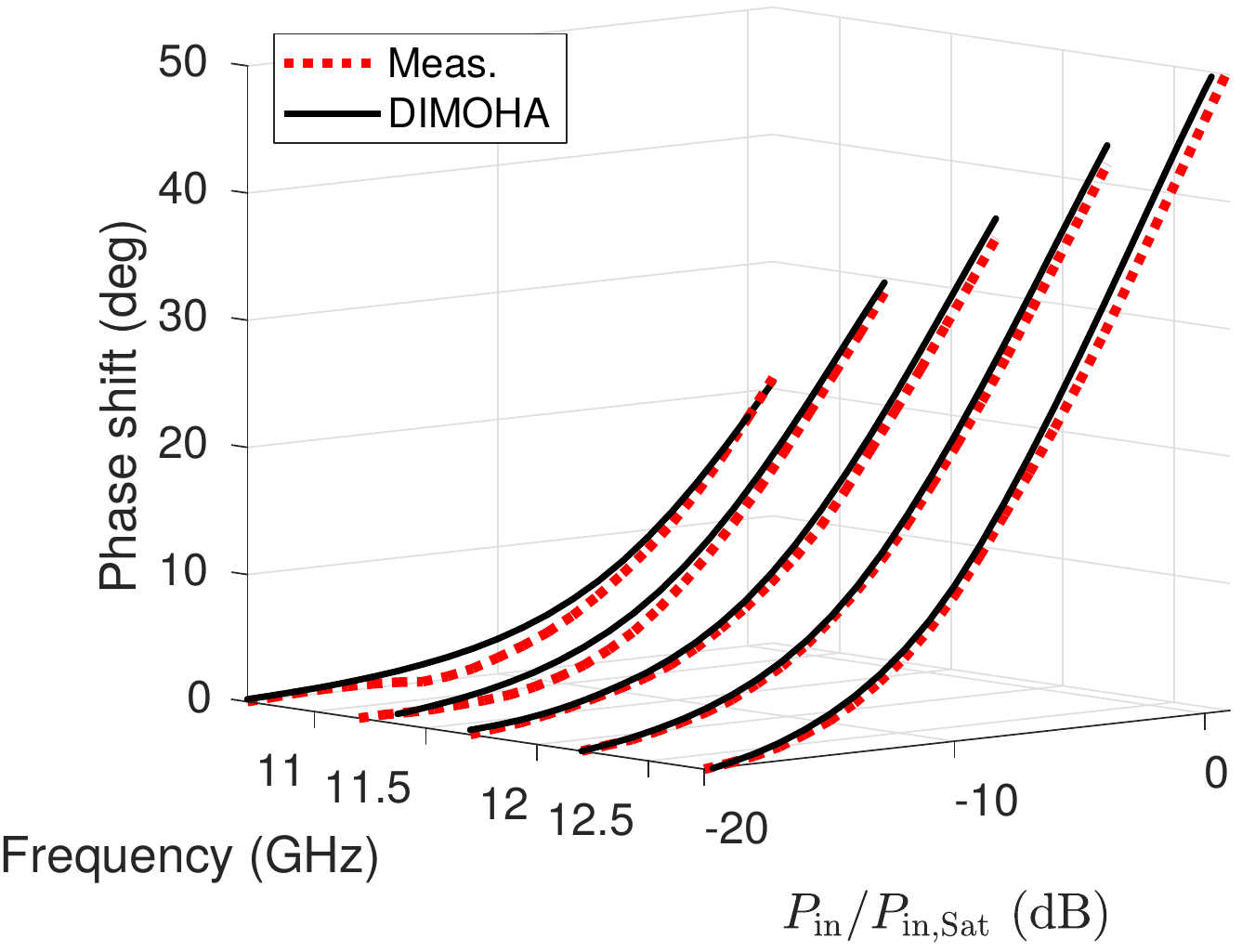} 
\caption{Comparison of \textsc{dimoha} (solid lines) and measurements from a Ku-band tapered helix TWT (dashed lines). a) Output power (AM/AM) depending on frequency and input power. b) Nonlinear phase shift (AM/PM) depending on frequency and input power.  Our attenuator model is set for the frequency $F =$ 11.7 GHz.}
\label{f:fig1AMAMPM}
\end{figure}

%

\subsection{Eigenfields}

Before computing time-dependent amplitudes, we also need the ``shape'' basis fields $\bsE_{n}, \bsH_{n},  \bsA_{n}$. In one dimension, the system is projected along the longitudinal $z$-axis and we can assume $\sH_{z,n} (z) = 0$. 
 For each cell, the on-axis shape fields $\sE_{z,n}$ for the electric field and $\sA_{z,n}$ for the vector potential are obtained as $\sE_{z,n} (z) = (2 \pi)^{-1} \int^{\pi}_{-\pi} \sE_{z, \beta} (z) \, \rme^{-\rmi n \beta d} \, \rmd (\beta d)$, from their discrete Fourier transforms (or inverse Gel'fand $\beta$-transform) $\sE_{z, \beta}$ and $\sA_{z, \beta} = - \rmi \sE_{z, \beta} / \omega$ \cite{min18}.
From the electromagnetic power, those eigenfields are computed as \cite{rys07,min18}
\begin{equation}
\sE_{z, \beta} (z) = \rme^{-\rmi \beta z}  \beta \, \sqrt{\frac{1}{d} \omega (\beta)  Z_{\rm c} (\beta) v_{\rm g} (\beta)} \, , \label{e:Ebeta1D}
\end{equation}
where $\omega$ the angular frequency, $v_{\rm g} = \derivep{\omega}{\beta}$ the group velocity and $Z_{\rm c}$ the impedance. 
To take into account any pitch tapering, eigenfields are recomputed at each different cell.
For TWTs, shape fields $\sE_{z,n},\sA_{z,n}$ are similar to cardinal sine functions centered at the $n$\textsuperscript{th} cell position \cite{ber11}.
In particular, $\sE_{z,n}(z), \sA_{z,n}(z)$ are computed for each points of the mesh $z_{\rm field}$ along the whole structure.
Those functions are needed to express the coupling with particles in the time integrator.

\subsection{Beam and space charge}

Since RF shape functions $\sA_{z,n}$ (responsible for the coupling with particles) are continuous and smooth, we can ease the computation by aggregating particles as macro-electrons.
In our simulations, each macro-electron (indexed by $k$) is characterized by its position $q_k(t)$ and velocity $v_k$ and contributes a source $-\sum^{N_{\rm e}}_k \rmi e v_k \sA_{z,n}(q_k)$ to \eqref{e:evo1}.
Their electric charge is $e = - I_0 \delta /v_0$, with $\delta$ their initial spacing and $I_0$ the cathode current.
Even if the relativistic correction is small, we build the model and algorithm with relativistic variables.
The initial beam velocity is $v_0 = c (1 - (1 - e_{\rm e} V_0/(m_{\rm e} c^2))^{-2})^{1/2}$, with $V_0$ the cathode potential, $e_{\rm e}$ the electron charge, $m_{\rm e}$ the electron mass, assuming the initial beam to be a line of equally spaced particles.
The mass of macro-electrons is $m =  m_{\rm e} \, e / e_{\rm e}$. 
We can also modify the velocity of each macro-electron to generate pulsed emissions or distribution functions like a gaussian beam.

As the discrete model decomposes only the circuit part of fields, it misses the space charge field. 
For $N_{\rm e}$ macro-electrons, the axial space-charge field in one dimension $E_{\rm sc}(q_{k}) = - \sum^{N_{\rm e}}_{k' \neq k} \partial_{z} \phi_{\rm sc} \left( q_{k}-q_{k'} \right)$ follows Rowe's model \cite{row65,ber11}
\begin{equation}
E_{\rm sc}(q_{k}) = \sum^{N_{\rm e}}_{k' \neq k} K \exp \left( - \left| \frac{Q_{k}-Q_{k'}}{b/2} \right| \right){\rm sign}  \left( \frac{Q_{k}-Q_{k'}}{b/2} \right) \, , 
\end{equation}
with $K = 2 e / (\pi \epsilon_0 b^2)<0$, 
for a cylindrical beam with radius $b$ (effects of the helix radius are neglected \cite{row65}), where $Q_k$ is the mesh point nearest to $q_k(t)$ on the mesh $z_{\rm field}$. The space charge field $E_{\rm sc}$ is pre-computed before the interaction (the charges and masses of particles are constants). After each displacement of macro-electrons, the space charge force is applied on them with this nearest-meshpoint approximation, allowing to drastically reduce the computational time.

 \section{Time integrator and diagnostics}
 \label{s:IntDiag}
 
The second and main part of \textsc{dimoha} is the numerical time integrator. 
The evolution of our system rests on a total hamiltonian \cite{jac99} initially composed of the electromagnetic term $\int_V (\epsilon_0 |\bfE|^2 + \mu_0 |\bfH|^2)/2 \, {\rm d} V$, the particles term $\sum_k \gamma_k m c^2$ (for macro-electrons), with $\gamma_k$ the Lorentz factor, and the space charge term $\sum^{N_{\rm e}}_{k' \neq k} \phi_{\rm sc} \left( q_{k}-q_{k'} \right)/2$.
This hamiltonian is re-expressed \cite{min18} with the discrete model and projected in 1D along the longitudinal axis. 
Associated Hamilton's equations
\begin{align}
\frac{{\rm d} \sV_{n}}{{\rm d} t} &= - \sum_{n'} \Omega_{n-n'} \sI_{n'}(t)  - \sum^{N_{\rm e}}_{k=1}  \rmi e \,  \dot{q}_k \, \sA_{z,n} (q_k)\, , \label{e:evo1bis} \\
\frac{{\rm d} \sI_{n}}{{\rm d} t}&=  \sum_{n'} \Omega_{n-n'} \sV_{n'}(t) \, , \label{e:evo2bi} \\
v_k &= \frac{1}{\gamma_k m} \left( p_k - e A_z(q_k) \right) \, , \\
\dot{p}_k &= \frac{1}{\gamma_k m} \left( p_k - e A_z \right) \derivep{A_z}{z}(q_k) + e E_{\rm sc}(q_k) \, , \label{e:evo4}
\end{align}
are the evolution equations of our sytem and are solved with our algorithm.
The dynamical variables become the amplitudes $\sV_{n} (t), \sI_{n}(t)$ of fields and the 1D positions $q_k(t)$ and velocities $v_k(t)$ of macro-electrons.
This dynamics is consistent with Maxwell equations and Lorentz force.

Our numerical integrator solves the evolution system from our $N$-body self-consistent hamiltonian model \cite{min18} and provides $\sV_{n} (t), \sI_{n}(t), q_k(t)$ and $v_k(t)$.
This evolution is solved from time $t=0$ to $t_{\rm max}$ with a given step size $\Delta t$. Typically, $t_{\rm max}$ is set to reach the stabilisation of the signal inside the tube.
In continuous waveform (CW) regime, the wave is excited (at prime cell $n=1$) at the frequency $F$ by setting $\sV_{n=1}(t) = U \cos (2 \pi F t)$, with $U$ depending on the input power.
To simulate multi-carriers, we simply set $\sV_{n=1}(t) = \sum_\chi U_\chi \cos (2 \pi F_\chi t)$, with the wanted frequencies and powers. A major advantage of \textsc{dimoha} is that we can inject any wave of interest, e.g.\ telecom signals with phase shift keying.

From our $N$-body self-consistent hamiltonian model \cite{min18}, we built a symplectic  integrator solving \eqref{e:evo1bis}-\eqref{e:evo4}. 
A symplectic transformation \cite{arn89}, usually associated with Hamilton's principle, preserves areas of the phase space. For numerical simulations, it allows \cite{hlw10} the increase of step size $\Delta t$ without inducing too large errors. 
The algorithm is parallelised to employ multi-processors. 
This parallelisation is done by dividing spatially the tube between the allocated threads,
associating each to a spatial domain with corresponding $z^{\rm thread}_{\rm cell}$ and $z^{\rm thread}_{\rm field}$. 
During the time evolution, those threads compute amplitudes $\sV_{n} (t), \sI_{n}(t)$ for $n \in z^{\rm thread}_{\rm cell}$ and particle motions (coupling + space charge effects) for $Q_k \in z^{\rm thread}_{\rm field}$.


The last part of \textsc{dimoha} computes the output power, the frequency spectrum of the output field and the distribution functions (velocities, positions) of particles.
The electromagnetic power (used in the next section) for a given frequency $F$ is \cite{pie50} $\langle P_{z} \rangle = |\hat{E_z}|^2 / (2 \beta^2 Z_{\mathrm{c}})$, with $\hat{E_z}$ the time-Fourier transform of Eq.~\eqref{e:FieldE} over a period $1/F$.

\begin{figure}[!t]
\centering
\includegraphics[width=\columnwidth]{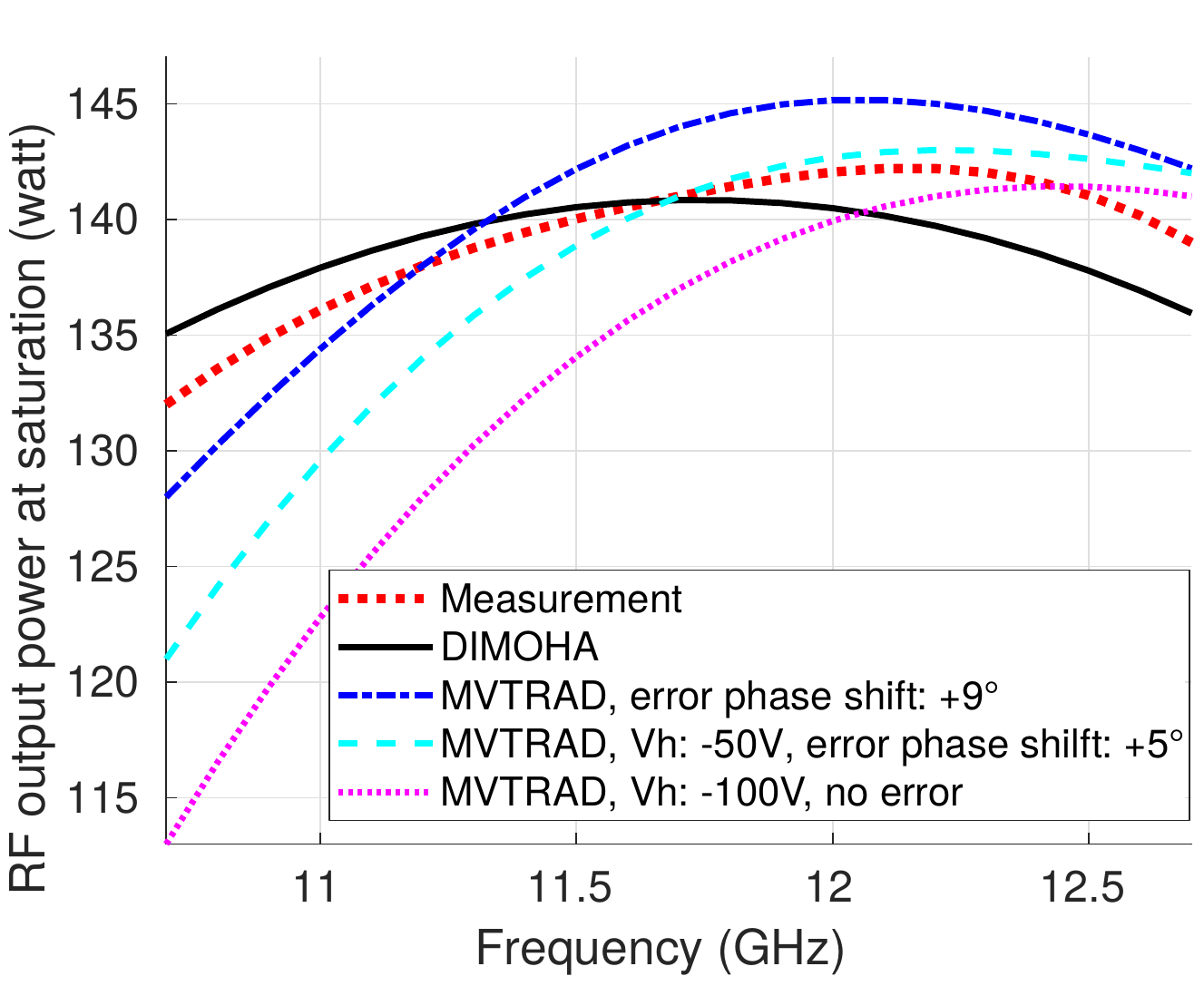}
\caption{Comparison of \textsc{dimoha} (solid black line), \textsc{mvtrad} (dashed lines) and measurements (dotted red line) from a 140W Ku-band tapered helix TWT  \cite{min19ivec}. Output power at saturation versus frequency. The cathode potential $V_0$ of \textsc{mvtrad} is shifted by a small $V_h$ (between 0 to $-100$ V) to adjust either the power output or the phase shift but not both at the same time. The phase shift of \textsc{dimoha} is identical to measurements (see Fig.~\ref{f:fig1AMAMPM}).}
\label{f:fig2MVTRAD}
\end{figure}

\begin{figure}[!t]
\centering
\includegraphics[width=\columnwidth]{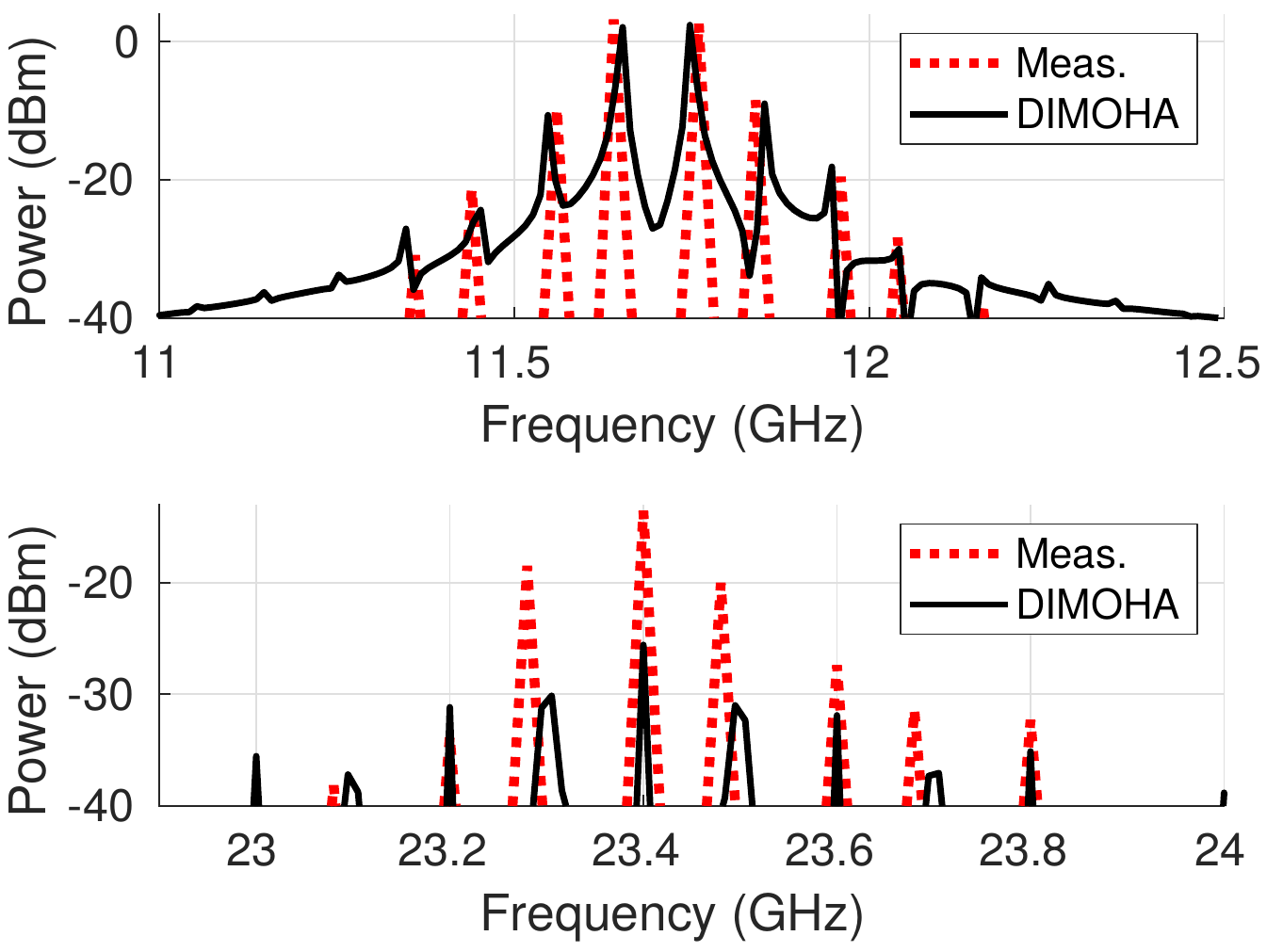}
\caption{Comparison of \textsc{dimoha} (solid black lines) and measurements (dotted red lines) from a 140W Ku-band tapered helix TWT \cite{min19ivec}. Output spectrum in two-carriers regime of \textsc{dimoha} (after $t_{\rm max} = 100$ ns) and the TWT. Input: $F_1 = 11.65$~GHz and $F_2 = 11.75$~GHz at $P_{{\rm in}} = -10$ dBm. Upper plot: fundamental signal. Lower plot: second harmonic.}
\label{f:fig2Multi}
\end{figure}

\section{Numerical results with a helix TWT}

In this section, we perform several comparisons between \textsc{dimoha} and measurements from a commercial 140W Ku-band pitch tapered helix TWT.

\subsection{Power and phase shift} 

We first compare \textsc{dimoha} with the output power and phase shift from a commercial 140W Ku-band tapered helix TWT \cite{min19ivec}.
Figure~\ref{f:fig1AMAMPM} displays the RF output power (AM/AM) and the nonlinear phase shift (AM/PM) at various frequencies. 
The phase shift in this context is the difference of phase between the input power $P_{\rm in}$ of interest and the input power at $-20$ dBm under the saturation $P_{\rm in,sat}$.
The horizontal axis reads $10\log_{10}(P_{\rm in} / P_{\rm in,sat})$ in dB. 
Small variations between \textsc{dimoha} and measurements may result from our incomplete management of tube defects (like hot voltage standing-wave ratio VSWR, defect in the beam confinement, and 3D asymmetry), from the beam model or from the violation of periodicity.

With initial spacing $\delta = 10^{-6}$ m, the model involves approximately 150~000 macro-electrons, leading to a total of 150~000 + 200 dofs. We set $t_{\rm max} = 5$~ns with step size $\Delta t=334$~fs.
The algorithm takes less than 10 min to reach steady state on a small desktop computer (8 threads with one Intel\textregistered Xeon\textregistered{} E5640 at 2.66GHz) and 2 min (32 threads with one Intel\textregistered Xeon\textregistered{} Gold 6142 at 2.6GHz) on a cluster \cite{meso}, with $\delta$ and $\Delta t$ small enough to ensure the convergence of the power to $\pm 0.1$~W. The parallelisation ensures a good scaling up to 8 processors.
This is a significant improvement over the 12~hours needed \cite{saf18} with \textsc{cst} for a similar tube.

\subsection{Comparison against a frequency code} 

Figure~\ref{f:fig2MVTRAD} displays the saturated output power over the complete tunable range of frequencies for the same Ku-band TWT and shows the comparison with \textsc{mvtrad} (an industrial standard for TWT design).
As shown, the cathode potential $V_0$ of \textsc{mvtrad} is shifted (between 0 to $-100$ V) to adjust either the power output or the phase shift but not both at the same time.
\textsc{dimoha} is almost overlapping with measurements, showing a better accuracy than \textsc{mvtrad}. Ongoing investigations with different TWTs support our claim that \textsc{dimoha} is more accurate than \textsc{mvtrad}.
Again, we surmise that small variations of \textsc{dimoha} versus measurements stem from the management of tube defects.

%

\subsection{Multiple carriers} 
\textsc{dimoha}'s prime advantage is that it considers the complete fields in time (not just wave envelopes) as well as the electron dynamics (positions and velocities). 
This enables our algorithm to work in nonlinear regime (trapping, chaos) and to inject any wanted type of wave, like multiple carriers and telecommunication signals. This is achieved by simply modifying $\sV_{n=1}(t)$ at the input cell.

We consider a simple two-carrier regime, setting $\sV_{n=1}(t) = U \cos (2 \pi F_1 t) + U \cos (2 \pi F_2 t)$, with $F_1 = 11.65$~GHz and $F_2 = 11.75$~GHz, and with the adequate $U$ ensuring $P_{{\rm in}}$ to be $-10$ dBm.
Figure~\ref{f:fig2Multi} displays the frequency spectrum at the same Ku-band TWT output.
The power at each frequency is computed from the time-Fourier transform of Eq.~\eqref{e:FieldE} over $t_{\rm max} = 100$ ns (minus $5$ ns to reach steady state).
All intermodulation products are reproduced near the two original carrier waves and their second harmonics.
For second harmonics (lower plot), we do not have an accurate calibration of the bench and this TWT is not adapted in this band.
The computation time is about 5 h for 100 ns  on a small desktop computer with the step size $\Delta t=83$~fs. 
Further investigations are in progress for multiple carrier and digital modulations.

\begin{figure}[!t]
\centering
\includegraphics[width=\columnwidth]{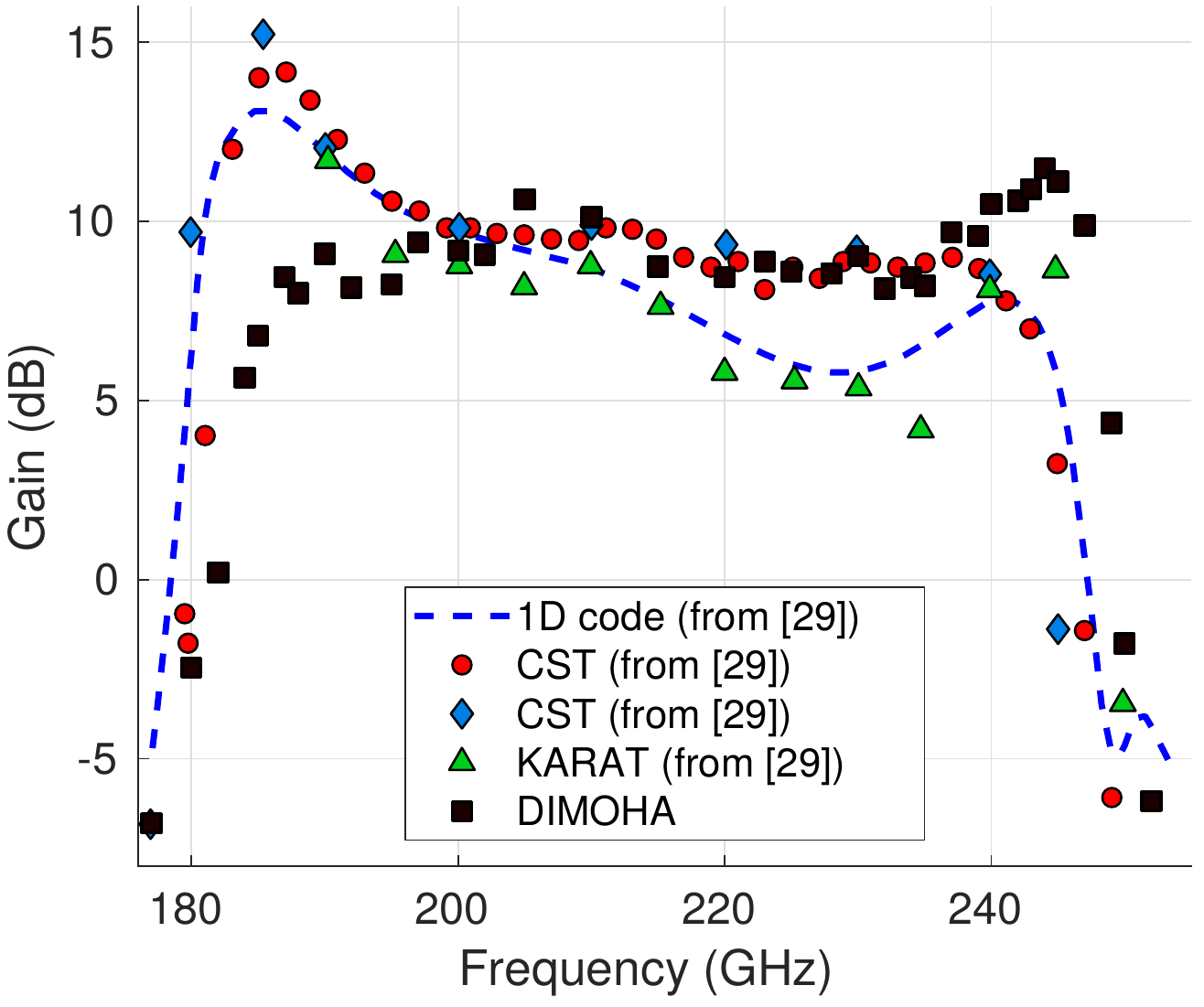}
\caption{Comparison of \textsc{dimoha} (squares), and simulations from Ref.\ \cite{kar18} using \textsc{cst} (circles and diamonds), \textsc{karat} (triangles) and their 1-D code (dashed blue lines), from a sub-THz folded waveguide TWT with a staggered double-grating SWS \cite{kar18} in small signal regime.
The agreement with \textsc{cst} in the middle of the band (200--240 GHz) is remarkable.}
\label{f:fig2Folded}
\end{figure}

\section{Folded waveguides} 
Since the theory is valid for any periodic structure, the algorithm is not limited to helix SWS but should handle all slow-wave structures.
We propose an example.
Ref.~\cite{kar18} provides inputs to model their sub-THz staggered double-grating TWT and includes results, used in Fig.~\ref{f:fig2Folded}, from 3D electromagnetic PIC codes \textsc{cst} \cite{cst} and \textsc{karat} \cite{karat}, and their 1D frequency model \cite{kar18}.
Figure~\ref{f:fig2Folded} displays the comparison between those approaches and \textsc{dimoha} in small signal regime. 
A truncation of coupling terms $\Omega_{n}$ to the nearest neighbour is enough.

The agreement with \textsc{cst} in the middle of the band (200--240 GHz) is great for a computation time about 23 min on a small desktop computer with the step size $\Delta t=18$~fs.
The band edge discrepancies may be due to various reasons. 
To compute the coupling terms $\Omega_{n}$, the discrete model requires the dispersion relation beyond the transmission band to be more accurate. 
Therefore, the data reading from the dispersion relation from Ref.~\cite{kar18} (limited from 177 to 260 GHz) is sensitive to any deviation, especially near the band edge. 
Moreover, space charge models are different (Ref.~\cite{kar18} considers a rectangular beam and we simulate a cylindrical beam).
However, experimental folded waveguides \cite{hu14} exhibit no peaks near edges.
The comparison on Fig.~\ref{f:fig2Folded} is not meant to explain those discrepancies nor to provide a full analysis of folded waveguides. 
It rather shows that \textsc{dimoha} is highly flexible \cite{meh} and it can handle various geometries in a fraction of the time required for PIC codes.

In addition, ongoing analyses are showing an excellent agreement between \textsc{dimoha} and \textsc{cst} for a W-band (92--95 GHz) folded waveguide TWT \cite{and17} with a sever.

\section{Conclusions and perspectives}

\textsc{dimoha} is a promising tool, for both industrial and research activities.
It also provides a new approach to analyse telecommunications signals for operators. Discrepancies exist between measurements and \textsc{dimoha}.
A more complete management of defects and a better space charge model could reduce this.
However, our algorithm is more accurate than the frequency domain code \textsc{mvtrad} and is greatly faster than classical time domain PIC codes.

Our algorithm bears five main advantages.
It can simulate complex signals like multiple carriers.
It is fast and robust.
It supports both helix and folded waveguide designs and it should be able to reproduce other kinds of SWS.
It reproduces harmonic generation, distortion phenomena and gain.
Finally, it can simulate nonlinear dynamics \cite{min18}.

We are investigating the cases of multiple carriers and digital modulations.
A more accurate version for folded waveguides and a 2.5D or 3D version of \textsc{dimoha} are under consideration.

\acknowledgments

The authors thank L{\'e}na{\"i}c Cou{\"e}del, Guillaume Fuhr, Magali Muraglia, Valentin Pigeon and Geoffroy Soubercaze-Pun for fruitful discussions, {\'E}lise Duverdier and Telma Pereira for their help  and anonymous reviewers for constructive comments. This work was granted access to the HPC resources of Aix-Marseille Université financed by the project Equip@Meso (ANR-10-EQPX-29-01) of the program “Investissements d’Avenir” supervised by the Agence Nationale de la Recherche.


\begin{thebibliography}{0}%
\makeatletter
\providecommand \@ifxundefined [1]{%
 \@ifx{#1\undefined}
}%
\providecommand \@ifnum [1]{%
 \ifnum #1\expandafter \@firstoftwo
 \else \expandafter \@secondoftwo
 \fi
}%
\providecommand \@ifx [1]{%
 \ifx #1\expandafter \@firstoftwo
 \else \expandafter \@secondoftwo
 \fi
}%
\providecommand \natexlab [1]{#1}%
\providecommand \enquote  [1]{``#1''}%
\providecommand \bibnamefont  [1]{#1}%
\providecommand \bibfnamefont [1]{#1}%
\providecommand \citenamefont [1]{#1}%
\providecommand \href@noop [0]{\@secondoftwo}%
\providecommand \href [0]{\begingroup \@sanitize@url \@href}%
\providecommand \@href[1]{\@@startlink{#1}\@@href}%
\providecommand \@@href[1]{\endgroup#1\@@endlink}%
\providecommand \@sanitize@url [0]{\catcode `\\12\catcode `\$12\catcode
  `\&12\catcode `\#12\catcode `\^12\catcode `\_12\catcode `\%12\relax}%
\providecommand \@@startlink[1]{}%
\providecommand \@@endlink[0]{}%
\providecommand \url  [0]{\begingroup\@sanitize@url \@url }%
\providecommand \@url [1]{\endgroup\@href {#1}{\urlprefix }}%
\providecommand \urlprefix  [0]{URL }%
\providecommand \Eprint [0]{\href }%
\providecommand \doibase [0]{http://dx.doi.org/}%
\providecommand \selectlanguage [0]{\@gobble}%
\providecommand \bibinfo  [0]{\@secondoftwo}%
\providecommand \bibfield  [0]{\@secondoftwo}%
\providecommand \translation [1]{[#1]}%
\providecommand \BibitemOpen [0]{}%
\providecommand \bibitemStop [0]{}%
\providecommand \bibitemNoStop [0]{.\EOS\space}%
\providecommand \EOS [0]{\spacefactor3000\relax}%
\providecommand \BibitemShut  [1]{\csname bibitem#1\endcsname}%
\let\auto@bib@innerbib\@empty
\end{thebibliography}%


\begin{thebibliography}{1}
%
\bibitem{min19epjh} D.~F.~G.~Minenna, F.~Andr{\'e}, Y.~Elskens, J-F. Auboin, F. Doveil, J.~Puech and {\'E}.~Duverdier, ``The Traveling-Wave Tube in the History of
Telecommunication,'' \textit{Eur. Phys. J. H}, vol. 44, no. 1, p. 1-36, 2019, doi: \href{http://dx.doi.org/10.1140/epjh/e2018-90023-1}{10.1140/epjh/e2018-90023-1}.
%
\bibitem{dov05} F. Doveil, D. F. Escande and A. Macor, 
``Experimental observation of nonlinear synchronization due to a single wave,''
\textit{Phys. Rev. Lett.} vol. 94, p. 085003, 2005, doi: \href{http://dx.doi.org/10.1103/PhysRevLett.94.085003}{10.1103/PhysRevLett.94.085003}.
%
\bibitem{saf18}
D. Safi, P. Birtel, S. Meyne and A. F. Jacob,
``A Traveling-Wave Tube Simulation Approach With CST Particle Studio,''
\emph{IEEE Trans. Electron Devices}, vol. 65, no. 6, p. 2257-2263, 2018, doi: \href{https://doi.org/10.1109/TED.2018.2798810}{10.1109/TED.2018.2798810}.
%
\bibitem{cst} \textit{CST Particle Studio}. Comput. Simulation Technol., Framingham, MA,
USA, 2017. [Online]. Available: \url{https://www.cst.com/products/cstps}.
%
\bibitem{karat} V. A. Tarakanov, \textit{User's Manual for Code KARAT}. Springfield, VA,
USA: Berkley Res., 1992.
%
\bibitem{wei97}
T. Weiland \textit{et al.}, ``MAFIA Version 4,'' \textit{AIP Conf. Proc.}, vol. 391, p. 65, 1997, doi: \href{http://dx.doi.org/10.1063/1.52369}{10.1063/1.52369}.
%
\bibitem{wal99}
P. Waller, ``Mod{\'e}lisation num{\'e}rique de l'interaction et diagnostic exp{\'e}rimental du faisceau d'{\'e}lectrons dans un tube {\`a} ondes progressives spatial,'' Ph.D. dissertation, Univ. Diderot-Paris 7, Paris, France, 1999.
%
\bibitem{ant97}
T. M. Antonsen, Jr., and B. Levush, ``CHRISTINE: A multifrequency parametric simulation code for traveling wave tube amplifiers,'' Naval Res. Lab Washington DC, Vacuum Electron. Branch, Washington, DC, USA, Tech. Rep. NRL/FR/6840–97-9845, 1997.
%
%
%
\bibitem{li09}
B. Li, Z. H. Yang, J. Q. Li, X. F. Zhu, T. Huang, Q. Hu, Y. L. Hu, L. Xu, J. J. Ma, L. Liao and L. Xiao, ``Theory and Design of Microwave-Tube Simulator Suite,''
\textit{IEEE Trans. Electron Devices}, vol. 56, no. 5, p. 919-927, 2009, doi: \href{http://dx.doi.org/10.1109/TED.2009.2015413}{10.1109/TED.2009.2015413}.
%
\bibitem{min17} D.~F.~G.~Minenna, Y.~Elskens and F.~Andr{\'e}, 
``Electron-wave momentum exchange and time domain simulations applied to traveling wave tube,'' in \textit{Proc. IEEE Int. Vacuum Electron. Conf. (IVEC)}, London, 2017, doi: \href{https://doi.org/10.1109/IVEC.2017.8289689}{10.1109/IVEC.2017.8289689}.
%
\bibitem{min19ivec} D.~F.~G.~Minenna, Y.~Elskens, F.~Andr{\'e}, J.~Puech, A.~Poy{\'e}, F.~Doveil and T.~Pereira, in \textit{Proc. IEEE Int. Vacuum Electron. Conf. (IVEC)}, Busan, 2019, doi: \href{https://doi.org/10.1109/IVEC.2019.8744984}{10.1109/IVEC.2019.8744984}.
%
\bibitem{min18} D.~F.~G.~Minenna, Y.~Elskens,  F.~Andr{\'e} and F.~Doveil, ``Electromagnetic power and momentum in N-body Hamiltonian approach to wave-particle dynamics in a periodic structure,'' 
\textit{EPL}, vol. 122, no. 4, p. 44002, 2018, doi: \href{http://dx.doi.org/10.1209/0295-5075/122/44002}{10.1209/0295-5075/122/44002}.
%
\bibitem{and13} F. Andr\'e, P. Bernardi, N. M. Ryskin, F. Doveil and Y. Elskens, ``Hamiltonian description of self-consistent wave-particle dynamics in a periodic structure,''
\textit{EPL}, vol. 103, no. 2, p. 28004, 2013, doi: \href{http://dx.doi.org/10.1209/0295-5075/103/28004}{10.1209/0295-5075/103/28004}.
%
\bibitem{arn89} 
V.~I.~Arnold, \textit{Mathematical Methods of Classical Mechanics}, Springer-Verlag, 1989.
%
\bibitem{hlw10}
E.~Hairer, C.~Lubich and G.~Wanner, \textit{Geometric numerical integration}, Springer, New York, 2010.
%
\bibitem{sai01}
I. Saitoh, Y. Suzuki, and N. Takahashi,
``The Symplectic Finite Difference Time Domain Method,''
\textit{IEEE Trans. Magn.}, vol. 37, no. 5, p. 3251-3254, 2001, doi: \href{http://dx.doi.org/10.1109/20.952588}{10.1109/20.952588}.
%
\bibitem{ber11}
P. Bernardi, F. Andr{\'e}, J.-F. David, A. Le Clair and F. Doveil, 
``Efficient Time-Domain Simulations of a Helix Traveling-Wave Tube,''
\textit{IEEE Trans. Electron Devices}, vol. 58, no. 6, p. 1761-1767, 2011, doi: \href{http://dx.doi.org/10.1109/TED.2011.2125793}{10.1109/TED.2011.2125793}.
%
\bibitem{ber11b}
P. Bernardi, F. Andr{\'e}, J.-F. David, A. Le Clair and F. Doveil, 
``Control of the Reflections at the Terminations of a Slow Wave Structure in the Nonstationary Discrete Theory of Excitation of a Periodic Waveguide,''
\textit{IEEE Trans. Electron Devices}, vol. 58, no. 11, p. 4093-4097, 2011, doi: \href{http://dx.doi.org/10.1109/TED.2011.2163410}{10.1109/TED.2011.2163410}.
%
\bibitem{rys09}
N.~M.~Ryskin, V. N. Titov and A. V. Yakovlev, 
``Nonstationary nonlinear discrete model of a coupled-cavity traveling-wave-tube amplifier,''
\emph{IEEE Trans. Electron Devices}, vol 56, no. 5, p. 928-934, 2009, doi: \href{http://dx.doi.org/10.1109/TED.2009.2016690}{10.1109/TED.2009.2016690}.
%
\bibitem{ter17}
A. G. Terentyuk, A. G. Rozhnev, and N. M. Ryskin,
``Discrete model of a folded-waveguide traveling-wave tube,''
in \textit{Proc. IEEE Int. Vacuum Electron. Conf. (IVEC)}, London, 2017, doi: \href{http://dx.doi.org/10.1109/IVEC.2017.8289668}{10.1109/IVEC.2017.8289668}.
%
\bibitem{min19pc} D.~F.~G.~Minenna, A.~G.~Terentyuk, Y.~Elskens, F.~Andr{\'e} and N.~M.~Ryskin, 
``Recent discrete model for small-signal analysis of traveling-wave tubes'' 
\textit{Phys. Scr.}, vol. 94, no. 5, p. 055601, 2019, doi: \href{https://doi.org/10.1088/1402-4896/ab060e}{10.1088/1402-4896/ab060e}.
%
\bibitem{kuz80} S. P. Kuznetsov,
``On one form of excitation equations of a periodic wave\-guide'',
{Sov. J. Commun. Technol. Electron.}, vol. {25}, p.~419-421, 1980.
%
\bibitem{rys07} N. M. Ryskin, V. N. Titov, and A. V. Yakovlev, 
Non-stationary nonlinear modeling of an electron beam interaction with a coupled cavity structure. I. Theory, in \textit{Modeling in Applied Electrodynamics and Electronics} (Saratov Univ. Press, Saratov, No. 8, 2007), p.~46-56.
%
\bibitem{rys} N. M. Ryskin and A. G. Terentyuk, private communication.
%
\bibitem{pie50} J. R. Pierce, \textit{Traveling Wave Tubes}, Van Nostrand, New York, 1950.
%
\bibitem{row65}
J. E. Rowe, \textit{Nonlinear Electron-Wave Interaction Phenomena}, San Francisco, Academic Press, 1965.
%
\bibitem{jac99} J.D. Jackson, \textit{Classical electrodynamics}, Wiley, New York, 3\textsuperscript{rd} ed, 1999.
%
\bibitem{meso} \textit{MesoCentre, Aix-Marseille Universit{\'e}}.  [Online]. Available: \url{https://mesocentre.univ-amu.fr/en/}
%
%
%
%
%
\bibitem{kar18}
T. A. Karetnikova, A. G. Rozhnev, N.~M.~Ryskin, A. E. Fedotov, S. V. Mishakin and N. S. Ginzburg,
``Gain Analysis of a 0.2-THz Traveling-Wave Tube With Sheet Electron Beam and Staggered Grating Slow Wave Structure,''
\emph{IEEE Trans. Electron Devices}, vol 65, no. 6, p. 2129-2134, 2018, doi: \href{http://dx.doi.org/10.1109/TED.2017.2787960}{10.1109/TED.2017.2787960}.



%




\bibitem{hu14}
Y. Hu, J. Feng,  J. Cai, X. Wu, Y. Du, J. Liu, J. Chen, and X. Zhang,
``Design and Experimental Study of a Widebandwidth
W-Band Folded Waveguide Continuous-Wave TWT,''
\textit{IEEE Trans. Plasma Sci.}, vol. 42, no. 10, p. 3380-3386, 2011, doi: \href{http://dx.doi.org/10.1109/TPS.2014.2350477}{10.1109/TPS.2014.2350477}.
%
\bibitem{meh} I. Mehinovi{\'c}, Private communication.
%
\bibitem{and17}
F. Andr{\'e}, S. Kohler, V. Krozer, Q. Trung Le, R. Letizia, C. Paoloni, A. Sabaawi, G. Ulisse and R. Zimmerman,
``Fabrication of W-band TWT for 5G small cells backhaul,''
in \textit{Proc. IEEE Int. Vacuum Electron. Conf. (IVEC)}, London, 2017, doi: \href{http://dx.doi.org/10.1109/IVEC.2017.8289653}{10.1109/IVEC.2017.8289653}.

\end{thebibliography}
\end{document}